\documentclass[conference]{IEEEtran}
\IEEEoverridecommandlockouts
\usepackage{cite}
\usepackage{amsmath,amssymb,amsfonts}
\usepackage{algorithm}
\usepackage[noend]{algpseudocode}
\usepackage{booktabs}
\usepackage{subcaption}
\usepackage{lipsum}
\usepackage{graphicx}
\usepackage{textcomp}
\usepackage{enumitem}
\usepackage{listings}
\usepackage[svgnames]{xcolor}
\makeatletter
\renewcommand{\ALG@beginalgorithmic}{\small}
\makeatother
\begin{document}
\title{Ingesting High-Velocity Streaming Graphs from Social Media Sources\\
\thanks{This work is partly funded by NSF Grant 1738411, and the AWESOME Project at the San Diego Supercomputer Center}
}

\author{\IEEEauthorblockN{Subhasis Dasgupta}
\IEEEauthorblockA{\textit{San Diego Supercomputer Center} \\
\textit{Univ. of California San Diego}\\
La Jolla, CA 92093, USA \\
sudasgupta@ucsd.edu}
\and
\IEEEauthorblockN{Aditya Bagchi}
\IEEEauthorblockA{\textit{Dept. of Computer Science}\\
\textit{RKMV Educational and Research Institute}\\
Howrah 711202, West Bengal, India\\
bagchi.aditya@gmail.com}
\and
\IEEEauthorblockN{Amarnath Gupta}
\IEEEauthorblockA{\textit{San Diego Supercomputer Center} \\
\textit{Univ. of California San Diego}\\
La Jolla, CA 92093, USA \\
a1gupta@ucsd.edu}
}

\maketitle

\begin{abstract}
Many data science applications like social network analysis use graphs as their primary form of data. However, acquiring graph-structured data from social media presents some interesting challenges. The first challenge is the high data velocity and bursty nature of the social media data. The second challenge is that the complex nature of the data makes the ingestion process expensive. If we want to store the streaming graph data in a graph database, we face a third challenge -- the database is very often unable to sustain the ingestion of high-velocity, high-burst data. We have developed an adaptive buffering mechanism and a graph compression technique that effectively mitigates the problem. A novel aspect of our method is that the adaptive buffering algorithm uses the data rate, the data content as well as the CPU resources of the database machine to determine an optimal data ingestion mechanism. We further show that an ingestion-time graph-compression strategy improves the efficiency of the data ingestion into the database. We have verified the efficacy of our ingestion optimization strategy through extensive experiments.
\end{abstract}

\section{Introduction}
\label{sec:intro}
A significant fraction of data used in Data Science today comes from streaming data sources. These include data from social media streams like Facebook and Twitter, IOT data from sensors, stock market data from stock exchanges and financial information sources. There are two broad categories of data science research for streaming data -- real-time analytics and non-real-time analytics. In the first case, analytics tasks can be performed on a small window of in-flight data as the data streams in. For example, Bifet et \cite{bifet2017extremely} develop a streaming decision tree technique that operates on an in-memory snapshot of data and adapts to changes in streams. In the second case, although the data is collected in real time, a data ingestion system needs to collect data for some time before the analytics operations can be executed. For example, computing hourly frequency distribution of hashtags from Twitter would require the system to store data because due to the high velocity of Twitter streams, an hour's worth of data will often exceed the memory capacity of the streaming system. In this case, the ingestion of the streaming data must keep up with the fluctuating data rates of the stream so that it does not have to resort to any load shedding scheme that were applied to a previous generation of stream processing systems (e.g., \cite{babcock2004load}). 

The reality, however, is that when the data gets more complex and needs pre-processing before storage, there a distinct \textit{bandwidth gap} between the data rate of the stream producer and the ingestion capability of the store that houses the streaming data. In this paper, we investigate the nature of and mitigation strategies for this bandwidth gap problem in the context of streaming social media data (JSON stream) that is transformed into a graph and stored in a graph database for analytics operations. We consider graph data to be more complex because, in contrast to relational records without strong integrity constraints, the nodes and edges of a graph are not independent of each other. Therefore, while edges of the graph arrive in random order in streaming data, the DBMS has to spend additional time to ensure that two neighboring edges ingested at different times from the stream are connected to the same node inside the DBMS. Thus, the ingestion cost of graph data is higher, leading to the bandwidth gap between the ingestion rate and the storage rate of streaming graphs. Interestingly, while processing high-volume graph data for network analysis is an emerging research area \cite{gui2019survey, samsi2018graphchallenge}, ingestion optimization of streaming graphs is still an unexplored area.

\begin{figure}[ht]
	\centering
	\includegraphics[width=0.48\textwidth]{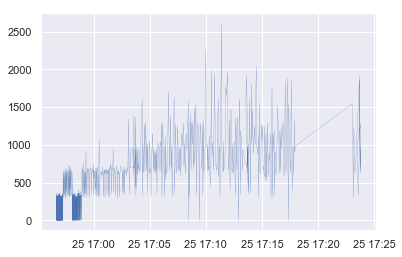}
	\caption{Performance Measurement During Direct Tweet Stream ingestion}
  \label{fig:burst}
\end{figure}

\begin{figure}[ht]
	\centering
	\includegraphics[width=0.48\textwidth]{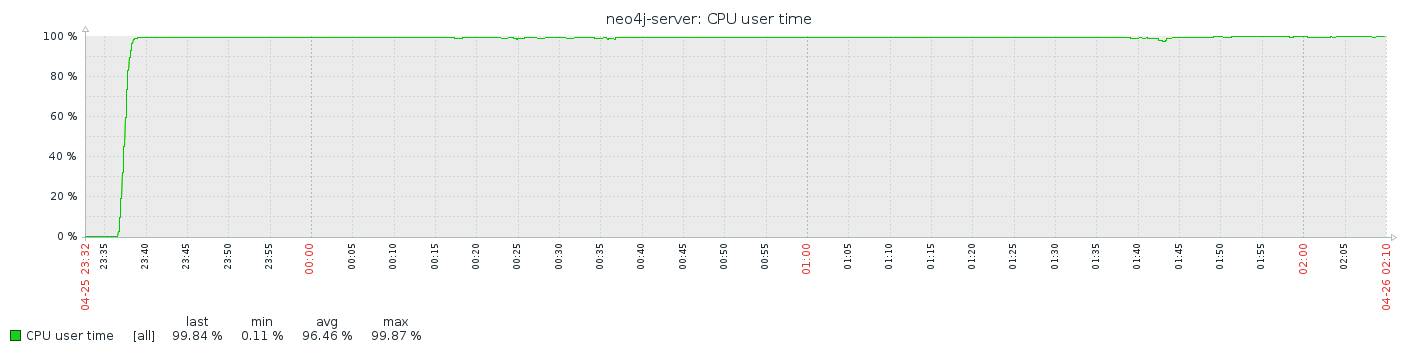}
	\caption{Performance Measurement During Direct Tweet Stream ingestion}
  \label{fig:cpu-user}
\end{figure}

\begin{figure}[ht]
	\centering
	\includegraphics[width=0.48\textwidth]{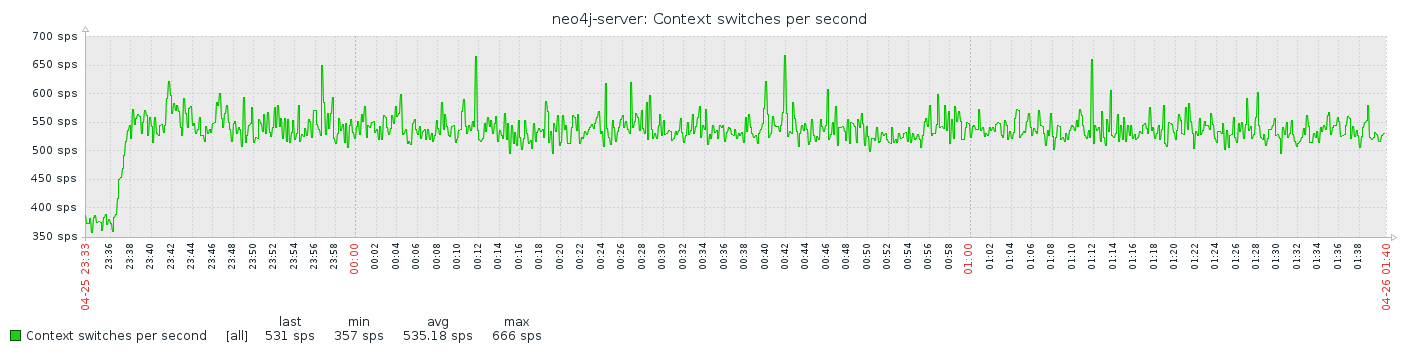}
	\caption{Performance Measurement During Direct Tweet Stream ingestion}
  \label{fig:context}
\end{figure}

\noindent \textbf{Example Use Case.} To motivate the problem, we present an illustrative use case from the domain of political science. The objective of the study is to understand patterns in political conversations and public opinions on Social Media in USA. One of the data sources for the study is Twitter; we use Twitter's streaming API (1\% sample) from which we filter tweets by using a set of domain-specific keywords. During politically charged times, the rate of Tweets received show a bursty behavior. Figure \ref{fig:burst} shows the rate of data arrival over a 25-second period. 
The figure shows the bursty nature of Tweet arrival and a peak value over 2500. This is in comparison with the average rate of 60 tweets/sec (1\% of 5787 tweets/minute \cite{twitterstat}) available in real-time using the Twitter API. This tweet stream is accepted by an \textit{ingestor process}, transformed into a graph model by a \textit{transformer process} and then pushed to a backend graph DBMS (Neo4J). As the velocity of tweets increases and the transformer process pushes increasingly more data to the store, the Neo4J machine reaches a 100\% user time (Figure \ref{fig:cpu-user}) while there is a small decrease in memory availability. The deterioration of system efficiency is further evidenced from the speed of context switching (Figure \ref{fig:context}). \textit{With no intervention, this results in a significant slowdown or a total failure of the DBMS server.}
This system failure points to a largely ignored aspect of the data science infrastructure -- with all the advances in improving ``Big Variety'' problems, ingestion of streaming graph data into graph databases has remained unaddressed \cite{calvillo2018management, reiner2018ingestion}.

In this paper, we address the above form of system failure by combining two  completely different approaches to the problem:
\begin{enumerate}[leftmargin=*]
    \item \textbf{Adaptive Buffering.} We develop an adaptive buffering scheme that monitors both the data arrival rate and the CPU load of the server and balances the effective ingestion load transmitted to the server.
    \item \textbf{Graph Compression.} We exploit the information redundancy in the social media data content to compress the graph load that would be ingested by the DBMS.
\end{enumerate}
For adaptive buffering, we create a predictive model of how the CPU load will be impacted by the buffer size and the variation of data content as the data rate fluctuates. We find that the buffer size itself is controlled by a metric related to \textit{the diversity of the data content} when the data is transformed to a graph. For graph compression, we make use of the observation that during a burst, a large number of users post about the correlated content and in the process, reuse hashtags created by others.

We show that using this combined approach, we can largely adapt to the velocity and burstiness, and only on rare occasions resort to spilling the incoming data to local storage of the ingestor machine.

\section{Ingesting Streaming Graphs}
\label{sec:arch}
Our stream processing engine has a pipelined architecture as shown in Figure \ref{fig:flow}. In the following, we first present the building blocks of this architecture and then present the controlling algorithms for managed stream ingestion. 
\subsection{Data Processing Pipeline}
\label{sec:dataproc}
\begin{figure}[ht!]
	\centering
    \includegraphics[width=0.55\textwidth]{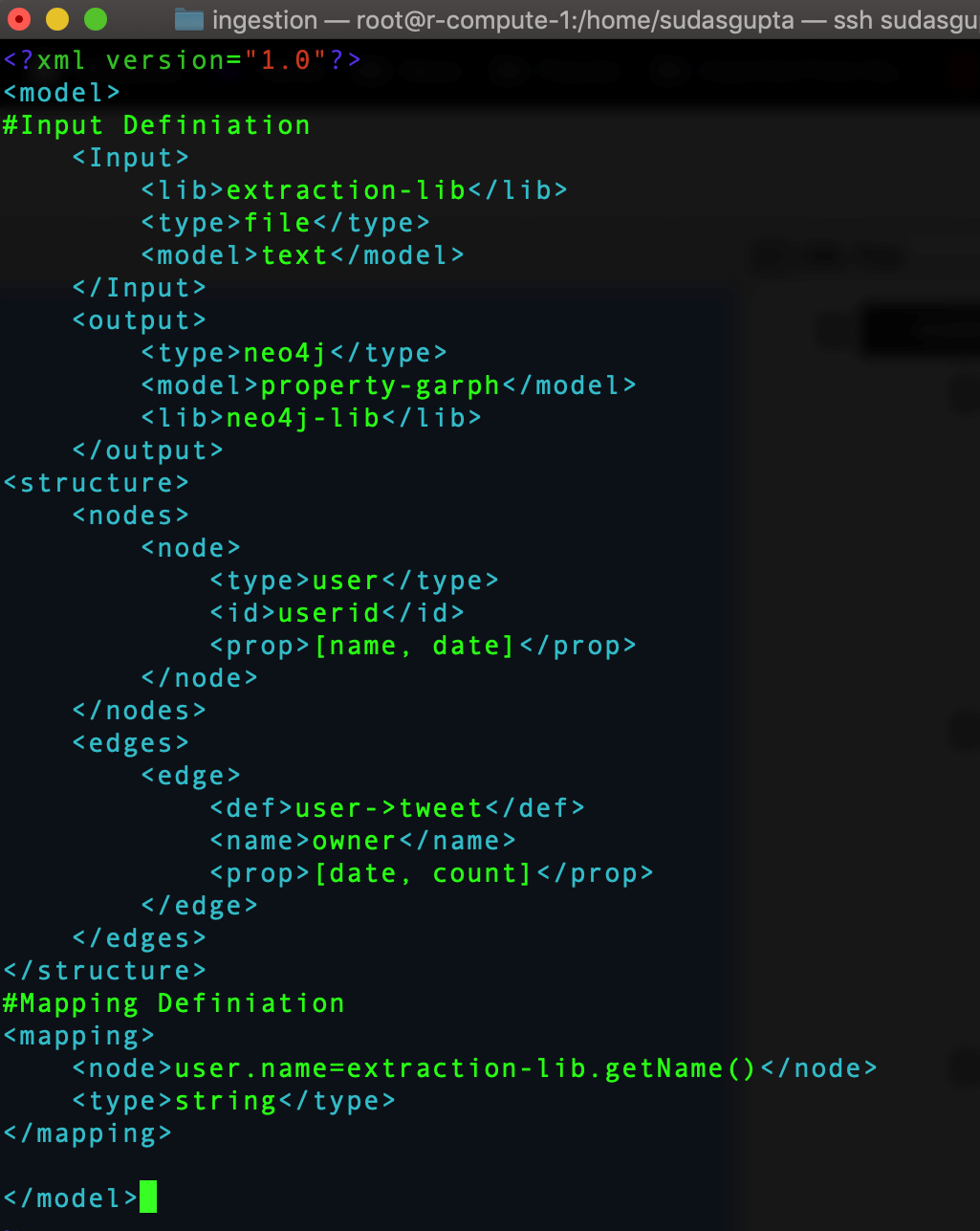}
	\caption{Basic Components of a XML Map file}
	\label{fig:xml-map}
\end{figure}

\begin{figure}[ht!]
	\centering
	\includegraphics[width=0.5\textwidth]{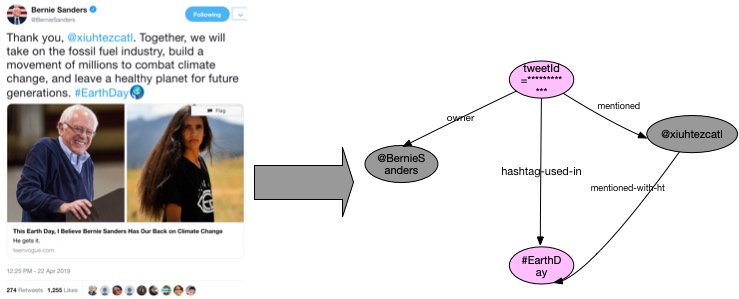}
	\caption{Model transformation for tweets}
	\label{fig:mdt-ex}
\end{figure}

The data processing pipeline, consisting of seven steps, is developed on top of a threaded, multiprocessing and partially distributed environment. The primary computation in the pipeline is data manipulation and transformation which is executed by breaking up the stream into mini-batches.
\begin{figure}[ht!]
	\centering
	\includegraphics[width=0.5\textwidth]{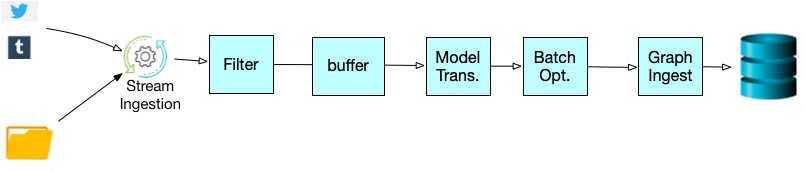}
	\caption{Ingestion Steps}
	\label{fig:flow}
\end{figure}
\begin{itemize}[leftmargin=*]
\item \textbf{Filter:} The ingestion process starts by filtering out data items (tweets) that do not satisfy the semantic requirements of the system. The filter is applied in two stages. The first set of filters is applied as a parameter of the streaming API provided by the data source (Twitter, in this case). In our example, we provide a set of keywords to the Twitter API for a specific application, and receive a data stream satisfying the filter. In the second phase, we apply a set of analysis-specific filtering criteria (e.g., remove tweets with only emojis). The choice of filtering criteria has a profound effect on the effective data rate of a stream. In our example, our keywords involve names of US politicians and some political issues. Therefore, whenever a political issue grabs public attention, we see a significant burst in the rate at which the data streams into our system. We have observed 15-45\% velocity fluctuations on a normal day and over 250\% fluctuation on extremely busy days. 
\item \textbf{Buffer:} The filtered data is collected in a buffer. As mentioned before, the size of the buffer is an important parameter in the efficiency of data ingestion. Using a buffer is a standard strategy, we determined that using a fixed buffer size does not effectively handle the problem of efficient ingestion management. 
In the case of a burst period, the CPU of the DBMS machine quickly goes to 100\% load and a large buffer is needed to absorb the bursting content and control the CPU load. However, using a large buffer also delays the ingestion process, because when the content of a large buffer is transmitted to the CPU, its ingestion load increases. 
To counter this dilemma, we have developed an adaptive buffer management strategy described in the next subsection that senses the impact of an upcoming burst and adaptively modifies the size of the buffer. We show that the factors impacting the required buffer size depends significantly on the data rate but also on the content of the data. The use of content in controlling the buffer size distinguishes our work from the traditional buffer management algorithms that only perform congestion control \cite{hirano1990traffic,vishwanath2009perspectives}.  
\item \textbf{Model Transformation: } Model transformation is the process of transforming data from its native form to a target form that conforms to the data model supported by the destination storage. In our case, tweets enter the system as a stream of tree-object (JSON) and needs to be converted into a property graph (a graph where both nodes and edges can have attributes and values).  In this step, a tweet (sometimes a tweet set, as shown later) is manipulated to construct typed nodes, labeled edges, node properties and edge properties. Figure \ref{fig:mdt-ex} shows an example of model transformation, where the JSON elements called \texttt{user} and \texttt{tweet} become nodes, but \texttt{hashtag}, a JSON property, is unnested and transformed into nodes because in the target graph model, a hashtag will be shared by a number of nodes.  The edges of the graph, namely \texttt{owner, mentioned, hashtag-used-in} and \texttt{mentioned-with-ht} are constructed from the JSON content. For instance the edge \texttt{mentioned-with-ht} connects a hashtag with a user who is mentioned in the tweet. In general, the model transformation uses a \textit{configuration file} that specifies the mapping between the input and target data. Fig. \ref{fig:xml-map} shows the node types and node properties the target graph must have, and a mapping section that specifies how these properties can be populated from the input (e.g., using the \texttt{getName}) function.
\item \textbf{Batch Optimizer:} The model-transformed data is prepared for ingestion by wrapping the data into INSERT clauses for the graph database. However, the actual insertion process is expensive because of the ingestion latency of the target DBMS. This is managed by grouping the INSERT operations into ``batches''. The batch optimization process determines the optimal batch size to improve system efficiency. Mini-batching  \cite{grover2015data,DBLP:conf/cidr/MeehanAZT17,DBLP:journals/corr/abs-1902-08271} is a standard strategy for optimizing throughput during ingestion. However, we exploit the observation that although the number of tweets increase during a burst, there is a high degree of redundancy among them. This presents an optimization opportunity because, the redundant portions of a graph must be ingested only once. In our case, this optimization takes the form of dynamic graph compression which utilizes the fact nodes like \texttt{user} and \texttt{hashtag} should be computed only once during batch creation. 
\item \textbf{Graph Ingestor: }  The graph ingestor has two parts. The first part manipulates the data structures of the model transformation step and constructs the ingestion instructions so that the graph can be ingested by the target DBMS. The construction implements the graph compression method. The second part is an interface between our pipeline (Fig. \ref{fig:flow}) and the graph DBMS. The ingestor pushes the data to the DBMS ingestion pool where the pool size is predefined and managed by the third party connectors. In our example, we choose the Neo4j DBMS, which uses \texttt{bolt} as a graph connector. 
\end{itemize}
\lstset{language=XML}

\begin{figure*}[ht!]
	\centering
    \includegraphics[width=0.7\textwidth, scale=.9]{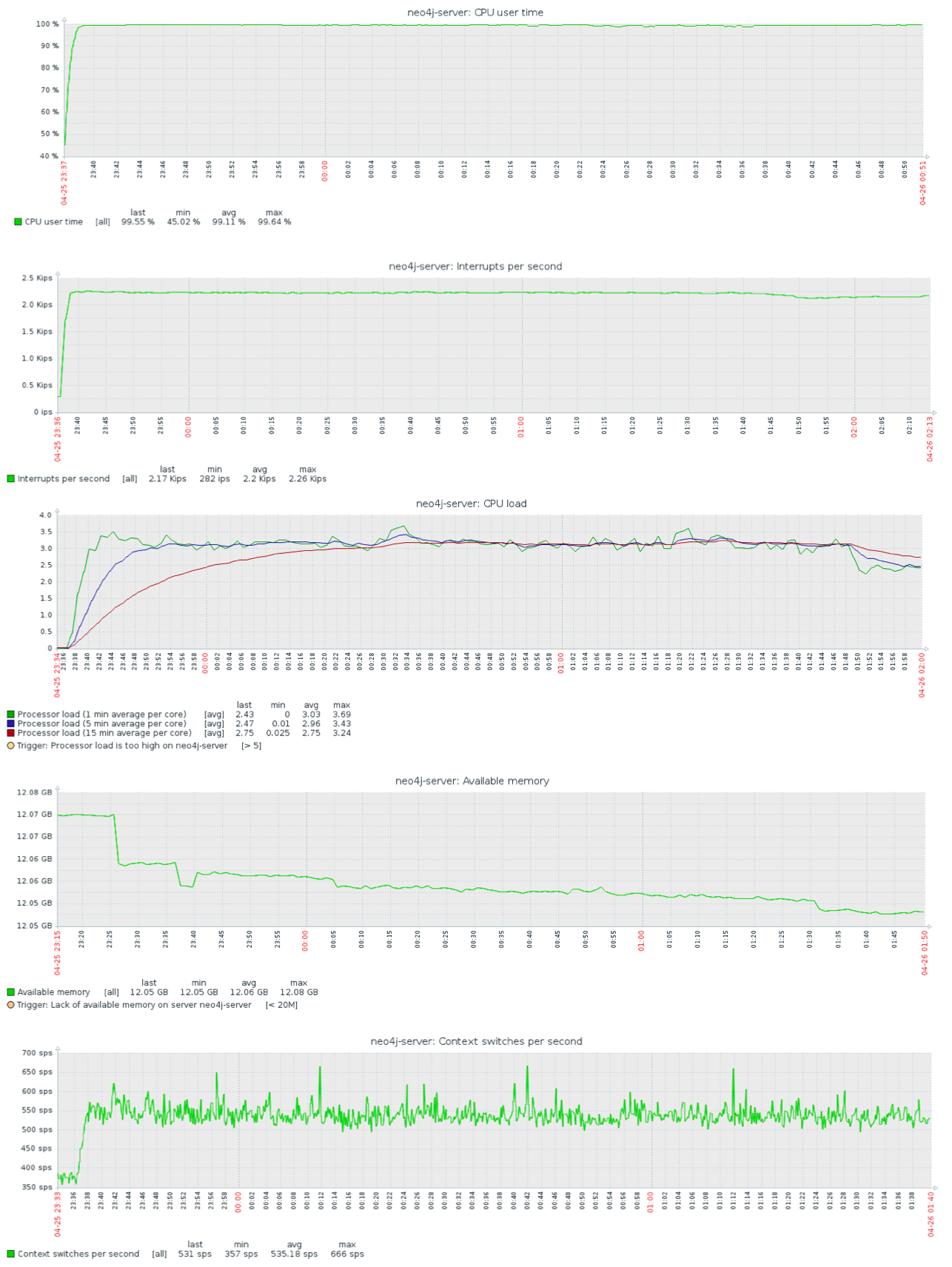}
	\caption{Effect of Uncontrolled Ingestion -- graph ingestion is a CPU-bound process}
	\label{fig:ing-wb}
\end{figure*}
\section{Ingestion Control}
In this section, we present the design principle behind the adaptive buffer control and graph compression algorithms to improve ingestion efficiency.    
\subsection{Modeling the Ingestion Problem}
\label{sec:adaptive}
Our strategy to manage the buffer for the streaming data, we first need to establish the factors that govern the buffer size. Let us first define a set of relevant parameters.
\begin{itemize}[leftmargin=*]
	\item \textbf{Graph Density($d$)} : The density of a graph $G(V,E)$ is the ratio of edges of $G$ to the maximum possible number of edges that can be induced by the nodes of $G$. Thus, density $d = 2(|E|)/(|V|*(|V|-1))$ where $|.|$ is the cardinality function.
	\item \textbf{Ingestion Buffer Size($\beta$) : } The ingestion buffer is the memory space used for all pre-processing operations on the streaming data, including filtering and model transformation. 
	\item \textbf{Effective Buffer/Output Buffer($\beta_e$) : } The effective buffer (or output buffer) is the buffer that contains the output of the model transformation. 
	\item \textbf{Bucket Diversity Ratio ($\rho$) :} A bucket $B[i]$ is a mini-batch of graph data that will be sent to the database for ingestion at time $i$. The bucket to be sent immediately has time index 0, and the bucket to be sent right after has index 1, and so forth. The diversity of a bucket is the proportion of new nodes (e.g., new hashtags) that appear in that bucket. The bucket diversity ratio $\rho$ is the average ratio of new nodes observed over $k$ temporal buckets. 
\end{itemize}
Based on the above parameters, the model to predict the effective buffer size $\beta_e$ is
\begin{equation} 
\beta_e = f(\rho, d)
\end{equation}
Notice that the model does not use $\beta$ as a variable because $\beta_e$ is generated from $\beta$. To set up the model, we assume that the model function $f$ does not depend on time, but the parameters need to be dynamically determined at each time chunk. We further assume that the effects of $\rho$ and $d$ on $\beta_e$ linearly separable. Thus, 
\begin{equation} \label{eq2}
\beta_e[i] = K[i].\phi_1(\rho[i]) + R[i].\phi_2(d[i])
\end{equation}
where the functions $\phi_1, \phi_2$ and their linear coefficients $K[i]$ and $R[i]$ need to be learned from the data. The result of the prediction model is presented in detail in Section \ref{sec:experiments}.
Once the parameters of this model are determined, we need to estimate how the buffer size i.e., the volume of data to be sent to the graph database for ingestion, impacts the stability of system resources on the DBMS side. The obvious candidate performance metrics to be considered on the DBMS side are memory, CPU user time (called CPU-usage later), context switching of the CPU, and interrupt per second. To simplify the model, we experimentally observe these metrics (Fig. \ref{fig:ing-wb}) over time, where no buffer control is exercised over input streams. A comparison of these performance metrics show that the CPU-usage rises from about 40\% to ~100\% in less than a second as a the number of ingested data records (i.e., the effective buffer size) increases. This effectively increases the delay time because the CPU spends longer updating the database.

The ingestion delay $I_{n}$ is the time gap between a record appearing at the stream, and the the record is ready for the query.  In other words, $I_n$ is the total time that the data stays inside the ingestion pipeline. There are two factors responsible for this delay --  buffer latency and system delay. Buffer latency refers to the time the time delay of a data item in the buffer due to the effective buffer size, while system delay refers to the delay that occurs because the CPU load was too high for the previous mini-batch, that results in a delay to get/process the next mini-batch. Let us assume that at any time point $i$ delay is the sum of bucket delay $\delta_{i}$ and system delay $\alpha_i$.  
Hence, the total system delay over $T$ time units is:
\begin{equation}\label{eq.3}
D =  \sum_{i=1}^{T} (\delta_{i} + \alpha_{i})
\end{equation}
If the expected value of CPU-usage at $i$th time-point is $\mu_{exp}[i]$  and the effective buffer size is $\beta_e[i]$. Then  
$\Delta \mu_{user}[n] = \mu_{user}[n] - \mu_{user}[n-1] $ is the change of CPU usage at $n$.
Since we intend to regulate the CPU use of the DBMS machine even when the streaming data is very large, our goal is to bound the value of $\Delta \mu_{user}[n]$ to achieve system stability. We observe that $\Delta \mu_{user}[n]$ monotonically increases with the increase of $\beta_e$, the effective buffer size of the ingestion control system increases. However, the nature of this monotonic function needs to be determined by a second predictive model, of the form
\begin{equation}\label{eq.4}
    \Delta \mu_{exp}[n] = f(\beta_e[n]) \Longleftrightarrow
        \mu_{exp}[n] = f(\beta_e[n]) - \mu_{exp}[n-1]
\end{equation}
Substituting Eq. \ref{eq2} from the first prediction model, we get
\begin{equation}\label{eq.5}
    \Delta \mu_{exp}[n] = f(K[n].\phi_1(\rho[n]) + R[n].\phi_2(d[n])) 
\end{equation}
In Section \ref{sec:experiments}, we experimentally estimate the form of the model and the parameters of these equations, and demonstrate that how we can effectively control the streaming ingestion problem.
\begin{figure}[ht!]
	\centering
    \includegraphics[width=0.45\textwidth]{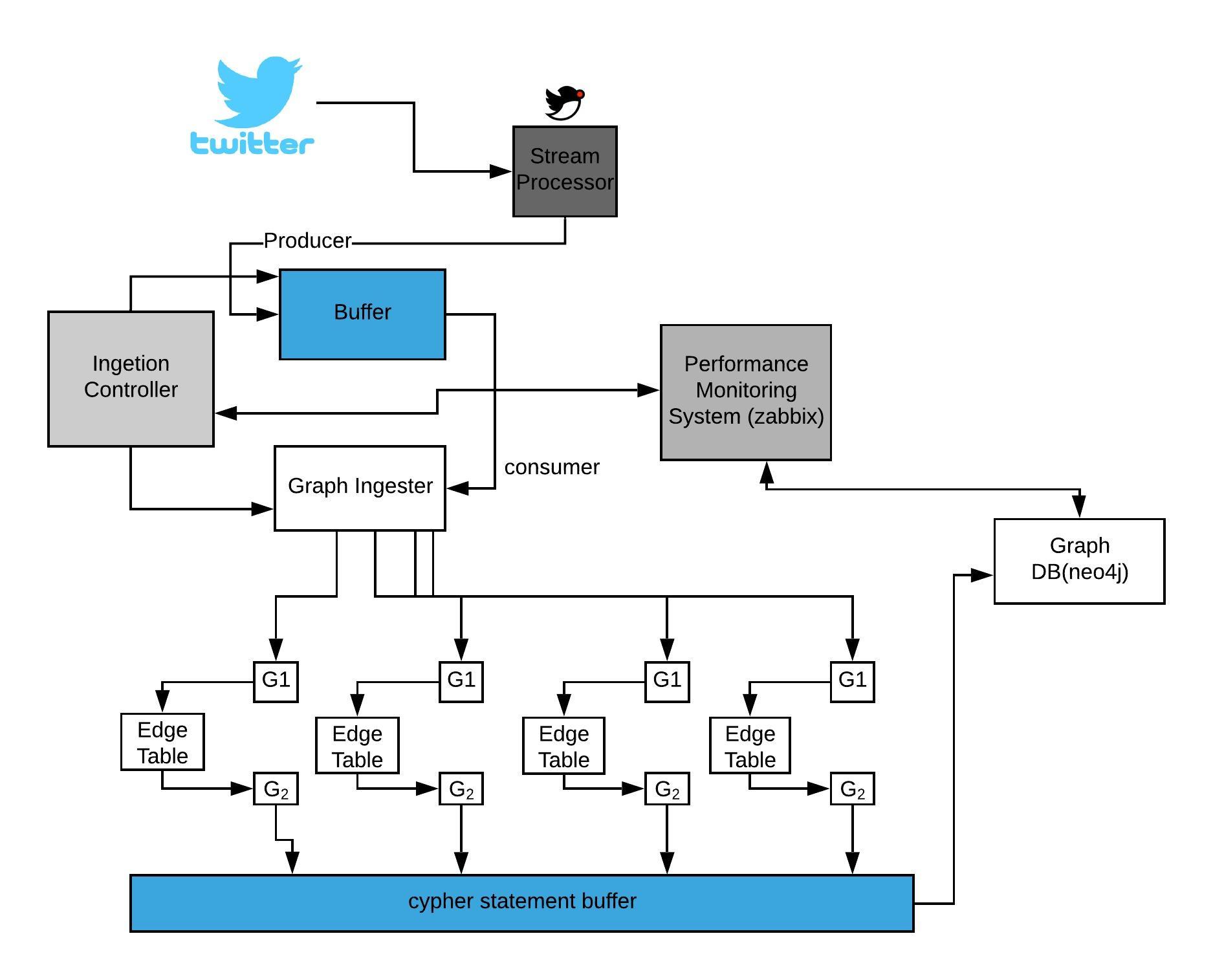}
	\caption{The Edge Table structure is implemented as a multithreaded structure. G1 represents the model transformation algorithm, and G2 is the graph insertion algorithm.}
	\label{fig:arch}
\end{figure}
\subsection{Algorithms}
In this section we present the algorithms referred to in previous subsections. The first algorithm \ref{alg:2} is the algorithm for model transformation (Section \ref{sec:dataproc}) where a JSON object is manipulated to construct a property graph. In the process, we implement the graph compression strategy mentioned in Section \ref{sec:intro}. The second algorithm implements the buffer control technique based on the prediction models from Section \ref{sec:adaptive}. The third algorithm controls the actual ingestion process that transmits data from the effective buffer to the DBMS server.
\begin{figure}[ht!]
	\centering
	\includegraphics[width=0.5\textwidth]{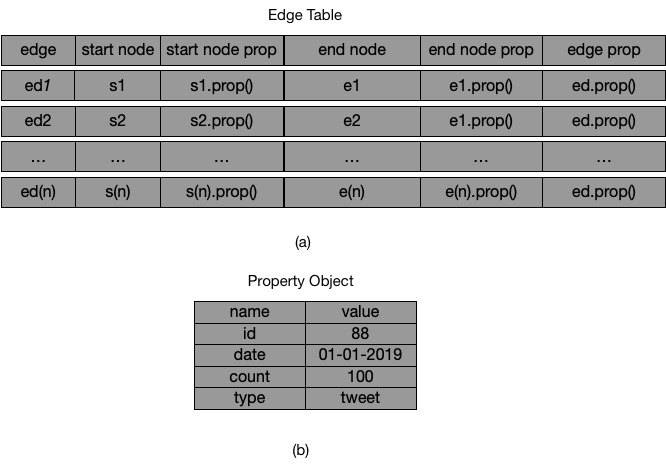}
	\caption{Edge Table Structure}
	\label{fig:edge-1}
\end{figure}
\newline
\textbf{Graph Model Transformation Algorithm: } Designed to be flexible, the graph model transformation algorithm tasks as input an XML-structured mapping file and a data extraction library which can parse an input data object and extract its sub-elements as needed. This extraction process depends on the data model of the input data file, and is not specific to any particular data set. In our case, the extraction library operates on any JSON file. The specificity of the problem-specific input and output is provided by the mapping file. This makes the translation more ``portable'' -- to choose a different data source (e.g., Reddit) that conforms to the same data model, one would only need to change the mapping file.
We choose an in-memory edge-centric data structure to represent the graph. The primary task of the algorithm is to extract information to populate an edge table and an indexed node list, followed by the insertion instructions from this in-memory representation of the property graph. Figure \ref{fig:edge-1} represents the structure of the edge table. Each edge has a unique id, start node, end node, start node properties, end node properties, and edge properties. Node and edge properties stored as a simple MAP object where the `key' is the name of the property and `value' is the property value. In addition, a set of table-level metadata like node density and diversity are computed for the edge table.
We use a special property `count' to handle duplicate edges. When we encounter a duplicate edge, we increase the value of the `count' (line  \ref{count-inc} of Algorithm \ref{alg:2}). The duplicate detection is handled by the procedure INSERTEDGE() (line \ref{insertedge} of Algorithm \ref{alg:2}). The algorithm keeps a node index to search nodes, and a list of connected nodes together.  The algorithm updates the node index in any insertion, while during the insertion it also searches for the duplicate edges. The indexed nodelist together with the deduplicated edge table are the necessary data structures used in the graph compression step described in Algorithm \ref{alg:3}. 

The \textsc{createedge}() algorithm works in batch mode. It takes a set of status or tweets, a map function, and extraction library as input. After initiation, the extraction function extracts nodes, node properties, edge properties (From line \ref{ce-1} to line \ref{ce-2}) and passes it to the \textsc{insertedge}() function. The run time complexity of the algorithm is linear in the number of edges. 
\begin{algorithm}[ht!]
	\caption{Graph Model Transformation Algorithm}\label{alg:2}
	\hspace*{\algorithmicindent} \texttt{Input: } List of records, record extraction library, XML mapping file \\
	\hspace*{\algorithmicindent} \texttt{Output: } edge table, node list, degree distribution, diversity ratio
	\begin{algorithmic}[1]
	\Procedure{CreateEdge}{$ListStatus[]$, datalib, map}
	  \State nodeList = new nodeList() \label{ce-1}
	  \State edgeList = new edgeTable() \Comment Initiating new edge table and node table
		\For {$status$ : $st[]$} 
		\State d = new datalib(map)\Comment Initialize a new data extraction functions
		\State edgeTypeList[] = d.getEdgeType()
		\For {edge : edgeTypeList}
		\State $stNode \leftarrow d.getStartNode(st)$ 
		\State $stProp[] \leftarrow d.getStartNodeProp(st)$ \Comment Returns map[name, value]
		\State $endNode \leftarrow d.getEndNode(st)$ 
		\State $endProp[] \leftarrow d.getEndNodeProp(st)$ \label{ce-2}
	  	\State \textsc{INSERTEDGE}(stNode, endNode, edgeTable, nodeIndex)
		\EndFor
		\EndFor
		\Return edgeTable
		\EndProcedure
\Procedure{InsertEdge}{stNode, endNode} \label{insertedge}
	\If {nodeIndex(stNode) \& nodeIndex(endNode)}
	\If {!(nodeIndex(stNode).getEdge(endNode))}
	\State nodeIndex(stNode).addEdge(endNode)
	\State nodeIndex(endNode).addEdge(stNode)
	\State addEdgeTable(stNode, endNode)
	\ElsIf{(nodeIndex(stNode).getEdge(endNode))} 
	\State edgeProp.count = edgeProp.count + 1 \label{count-inc}
	\EndIf
	\State nodeIndex.add(stNode)
	\State nodeIndex.add(endNode)
	\EndIf
	\If {(!nodeIndex(stNode) ) \& nodeIndex(endNode)}
	\If {!(nodeIndex(endNode).getEdge(stNode))}
	\State nodeIndex(endNode).addEdge(stNode)
	\State addEdgeTable(stNode, endNode)
	\EndIf
	\State nodeIndex.add(stNode)
	\EndIf
	\If {(nodeIndex(stNode) ) \& !nodeIndex(endNode)}
	\If {!(nodeIndex(stNode).getEdge(endNode))}
	\State nodeIndex(stNode).addEdge(endNode)
	\State addEdgeTable(stNode, endNode)
	\EndIf
	\State nodeIndex.add(endNode)
	\EndIf
	\If {(nodeIndex(stNode) ) \& nodeIndex(endNode)}
	\If {!(nodeIndex(stNode).getEdge(endNode))}
	\State nodeIndex(stNode).addEdge(endNode)
	\State addEdgeTable(stNode, endNode)
	\EndIf
	\If {!(nodeIndex(stNode).getEdge(endNode))}
	\State nodeIndex(stNode).addEdge(endNode)
	\State addEdgeTable(stNode, endNode)
	\EndIf
	\State nodeIndex.add(endNode)
	\State nodeIndex.add(endNode)
	\EndIf
	\Return edgeTable, nodeIndex
\EndProcedure
\end{algorithmic}
\end{algorithm}
\newline
\textbf{Buffer Controller Algorithm: } The objective of the buffer control algorithm (Algorithm \ref{alg:1}) is to improve the system stability during ingestion. Since graph ingestion is a CPU bound process, our algorithm maintains the `CPU-usage' (the CPU utilization percentage for the user space) level within acceptable bounds (called $cpu_{min}$ and $cpu_{max}$ respectively). Specifically, as the date rate fluctuates, this algorithm controls the CPU load by adjusting the buffer size, within the range [$\beta_{min}, \beta_{max}$].  The edge table computes diversity ratio, velocity, and the degree distribution of the nodes (line \ref{in-1} to \ref{in-3} ), and the Zabbix API supplies CPU-usage.  Hence, the input of the algorithm is average CPU-usage data and the edge table. Depending on the data velocity and the diversity for a particular time range,  we predict the actual buffer size by using multivariate linear regression. Next, we estimate the possible maximum buffer size from  CPU-usage data and the ``acceleration'', i.e., second derivative of data rate. We use linear regression to compute predicted CPU-usage. The steps of the buffer control algorithm are detailed as follows. 
\begin{enumerate}[leftmargin=*]
    \item With the input, the algorithm estimates effective buffer size, expected CPU-usage, and the `velocity', i.e., the first derivative of the data rate.
    \item If the expected CPU-usage is higher than $cpu_{max}$, increase the buffer size by $\theta_1$ (a constant in the range [0,1]) times the available memory. 
    \item It measures if the CPU-usage is $\theta_2$ times higher than $cpu_{max}$. If so, it writes to the local disk, which we call \textit{data throttling}. Here, $\theta_1$ and $\theta_2$ are system specific constants determined experimentally for our testbed. 
    \item If the expected CPU-usage is lower than $cpu_{max}$, we push the data to the graph database.
    \item While the buffer size is greater than $\beta_{min}$, it decreases the buffer size by $\theta_2$ times the available memory. This increase the availability of the data because with a lower buffer size, the ingestion latency is improved.
    \item If the CPU is $\theta_2$ times lower than $cpu_{max}$, it reads from the disk where the data was stored during throttling, and pushes it forward to the DBMS server. 
    \item At every step of the above process, the buffer size, expected CPU, and velocity are supplied by the PREFMON function, which uses our prediction models the CPU user time. 
\end{enumerate}
\begin{algorithm}[ht!]
	\caption{Buffer Control Algorithm}\label{alg:1}
	\hspace*{\algorithmicindent} \texttt{Input: } CPU performance Data($\mu$), EdgeTable($et$), cpu[max, min] , nodeIndex($n$), maximum and  buffer($\beta_{max}, \beta_{min}$)  \\
	\begin{algorithmic}[1]
	\Procedure{BufferControl}{$\mu, et, \mu[min, max], n$} 
	\State $\beta$,  $\mu_{exp}$, $s \leftarrow PerfMon(et, \mu, n) $
	    \If {cpu[max] $\leq$ $\mu_{exp}$}
	     \State echo ``CPU High Alert''
	     \State sleep($n$) \Comment Sleep $n$ time period $n$ depends on configuration
	     	\If {$(\beta + (\theta_2*\beta)) \leq \beta_{max}$}
	     	$\beta \leftarrow \beta + \theta_2*\beta$ \Comment Increase the size of buffer to delay the ingestion 
	     \EndIf 
	     \State $\beta_{exp}$,  $\mu_{exp}$, $s \leftarrow PerfMon(et, \mu, n) $
	     \If {$\theta_2*cpu[max]\leq \mu_{exp}\& s \geq 0$} 
	      \State FlushDataToDisk()\Comment throttle the data to disk assuming  $\theta_1$ is a configuration properties varies from system to system 
	     \EndIf
   	     \EndIf   	     
	  \If {  $\beta \geq \beta_{max}$ }
	  \If {  $\mu_{exp} \leq cpu[max]$ }
	   		\textsc{graphpush}($e$)
	   	\If {$(\beta - (\theta_2*\beta)) \geq \beta_{min} $}
	   	\State $\beta \leftarrow \beta - \theta_2*\beta$\Comment decrease the size of buffer to reduce ingestion time or latency 
	   	\EndIf 			   
	   	\If {$\theta_2*cpu[min]\geq$ $\mu_{exp}$} 
	        \State LoadDiskFromDisk(expBuf.Size())\Comment Writing Data to the Disk
	   \EndIf
	    \EndIf	    
	    \EndIf
	    \EndProcedure
	    \Procedure{PerfMon}   {edgeTable, $CPU_{user}$} 
	     \State $\delta$ = edgeTable.getDiversityRatio() \label{in-1}
	    \State $\nu$  = edgeTable.getNodeDensity()
	    \State $e$ = edgeTable.size() + nodeIndex.size() \label{in-2} \Comment the size of effective buffer $e$
	    \State $\beta_{exp}$ = $M*\nu^2$ + $N/\delta$ \label{in-3} \Comment Used machine learning model to derive the degree and coeff
	    \State $CPU_{exp} \leftarrow A* \beta_{exp} + B*avg(\mu_{user})$ \label{in-4} \Comment Used machine learning model 
	    \State $ s \leftarrow getCPUSlope()$ \Comment Simple regression to get slop from the data
	    \State Return $e$,  $CPU_{exp}$,$s$
	     \EndProcedure
	    \end{algorithmic}
\end{algorithm}
\textbf{Graph Insertion Algorithm:} The \textsc{graphpush} method used to transmit the graph from the ingestion machine to the graph database is explained in Algorithm \ref{alg:3}. The method converts the data from the edge table, node list and node properties to construct the insertion instruction using the \textsc{create} and \textsc{merge} statements of Cypher, the language our target database. 
Algorithm \ref{alg:3} creates node and edge ingestion statements in Cypher (From line number \ref{algo:3:node-ing-start} to  \ref{algo:3:node-ing-end} in \ref{alg:3})  by extracting start node and end node from the edges of the edge table. It uses an indexed list(line \ref{algo:3:list}) to ensure that nodes are created only once in the target database. For each commitment transaction, it also checks the integrity constraint that the nodes referred to in the edge table also exists in the node list. Since a commit to the DBMS may fail due to many practical reasons like network failure, the method stores data to the local memory until the timeout. It uses 3rd party data connectors to create a fault tolerant connection to the DBMS and to maintain a suitable data pool size at the DBMS.
\newline
 \textbf{Graph Compression:} In the previous process, at the end of the edge list traversal, the algorithm creates a set of unique node insertion instruction and the corresponding edge instructions as well. Our process guarantees the removal of the duplicate node in this stage, while it compresses the number of edges during the edge table creation process. Hence, our algorithm ensures compressed and minimum ingestion for each of the buffers. In Section \ref{sec:experiments} we have show how the compression reduces the ingestion load and demonstrate the interaction between the compression and the buffer size. 
\begin{algorithm}
	\caption{Graph Commit Algorithm}\label{alg:3}
	\hspace*{\algorithmicindent} \texttt{Input: } edgeTable, XML Configuration file
	\begin{algorithmic}[1]
		\Procedure{GaphPush}{$edgeTable,  map.xml$} 
		\If{$poolSize \leq maxPoolSize$}
		\State $db \leftarrow connect(map.xml)$\Comment Create Graph connection 
		\EndIf
		\State $nodeList[] \leftarrow \emptyset$\Comment Hash List for nodes \label{algo:3:list}
		\For {\{ i=0 ; $i< edgeIdList.size();$ i++ \}} 
		\If{ nodeList.contains(edgeList.getStartNode()) }\label{algo:3:node-ing-start}
		    \State nodeList.add(edgeList.getStartNode())
			\State $node[] \leftarrow createNode(edgeList.getStartNode())$
	    \EndIf
		\If{ node.contains(edgeList.getEndNode()) }
		\State nodeList.add(edgeList.getEndNode())
			\State $node[] \leftarrow createNode(edgeList.getEndNode())$
		\EndIf \label{algo:3:node-ing-end}
		\State $edge[] \leftarrow createEdge(edgeList.getEdge())$
		\EndFor
		\State $nodeList[] \leftarrow \emptyset$
		\If{commit = true}
		\State $db.execute(node)$
		\State $db.execute(edge)$
		
		\EndIf
		\If{commit = false}
		\State $archive.store(node, edge)$
		\EndIf
		\EndProcedure
	\end{algorithmic}
\end{algorithm}
\label{sec:architecture}
\section{Experiments}
\label{sec:experiments}
\begin{table*}
	\caption{Experimental results of the Coefficient and Error Estimations}
	\label{tab:regress}	
	\centering
	\begin{subtable}[t]{0.3\linewidth}
		\centering
		\vspace{0pt}
		\begin{tabular}{llll}		
		Max CPU & MAE   & \textbf{MSE} & RMSE  \\
		40      & 2.97  &   17.57     & 4.19  \\
		55      & 3.29 & 39.66        & 6.29  \\
		55      & 1.35  & 15.98        & 3.99 
	\end{tabular}
		\caption{$\mu_{exp} = A.\mu_{n-1} + B.log( \beta_{exp})$}
		\vspace{0.3cm}		
		\begin{tabular}{llll}
			Max CPU & MAE   & \textbf{MSE} & RMSE  \\
			40      & 2.79  & 17.27        & 4.19  \\
			50      & 3.09  & 37.19        & 3.09  \\
			55      & 2.26 & 33.44       & 5.78
		\end{tabular}
	\caption{$\mu_{exp} = A.\mu_{n-1} + B.\beta_{exp}^2$}
	\vspace{0.3cm}		
\begin{tabular}{llll}
	Max CPU & MAE   & \textbf{MSE} & RMSE  \\
	35      & 2.89  &   17.56     & 4.19  \\
	50      & 3.29  & 37.19      & 6.09 \\
	55      & 2.86 & 33.44        & 5.78  		
\end{tabular}
\caption{$\mu_{exp} = A.\mu_{n-1} + B.\beta_{exp}$}
	\end{subtable}\hfill
	\begin{subtable}[t]{0.3\linewidth}
		\centering
		\vspace{0pt}
				\begin{tabular}{llll}		
				Max CPU & MAE   & \textbf{MSE} & RMSE  \\
				40      & 2.92  &   17.57     & 4.19  \\
				50      & 3.29  & 37.19        & 6.09 \\
				55      & 2.26 & 33.44        & 5.78 
			\end{tabular}
		\caption{$\mu_{exp} = A.log(\mu_{n-1}) + B.log(\beta_{exp})$}
		\vspace{0.3cm}		
		\begin{tabular}{llll}
			Max CPU & MAE   & \textbf{MSE} & RMSE  \\
			40      & 2.79  & 17.27        & 4.19  \\
			55      & 3.09  & 37.19        & 6.09 \\ 
			50      & 2.26 & 33.44       & 5.78 		
		\end{tabular}
	\caption{$\mu_{exp} = A.\mu_{n-1} + B.log(\beta_{exp})$}
	\vspace{0.3cm}		
	\begin{tabular}{llll}
		Max CPU & MAE   & \textbf{MSE} & RMSE  \\
		40      & 2.89  &   17.56     & 4.19  \\
		50      & 3.29  & 37.19      & 6.09 \\
		55      & 2.86 & 33.44        & 5.78  		
	\end{tabular}
		\caption{$\mu_{exp} = A.\mu_{n-1} ^2+ B.log(\beta_{exp})$}
		\label{tbl:two}
	\end{subtable}\hfill
\begin{subtable}[t]{0.3\linewidth}
	\centering
	\vspace{0pt}	
    	\vspace{0pt}
    \begin{tabular}{llll}
    	Param & 40   & \textbf{50} & 55  \\
    	A     & .009  &   .008     & 0.09 \\
    	B     & .001 & .0024      & .003 \\
    	Intercept      & 0.541 & 5.29        &  1.96 		
    \end{tabular}
    \caption{$\mu_{exp} = A.\mu_{n-1}+ B.log(\beta_{exp})$}
\end{subtable}
\end{table*}
\noindent \textbf{Environment and Deployment.}
\begin{figure}[t]
	\centering
	\includegraphics[width=0.45\textwidth, scale=0.3]{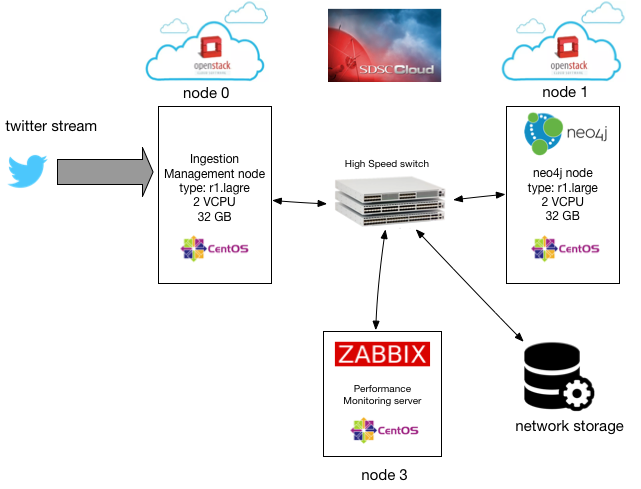}
	\caption{Deployment Diagram of our Test bed }
	\label{fig:depl-1}
\end{figure}
The experimental testbed for our work is a cloud computing environment (SDSC OpenStack cloud\footnote{https://www.sdsc.edu/services/ci/cloud.html}) that runs CentOS 7. The deployment diagram in Figure \ref{fig:depl-1} shows an ingestion server node, a database node with Neo4J 3.6 and a performance monitoring server (Zabbix 4.2 agent with json-rpc api support). Each node contains 2 VPUs and 32 GB memory and connected using high performance switches. The underlying processors are Intel Westmere (Nehalem-C) with 16384 KB Cache and around 2.2 GHz clock.\\
\textbf{Data Set.} The data set for our experiments come from the ``Political Data Analysis'' project at the San Diego Supercomputer Center. The project collects tweets continuously. In our experiments, we used two forms of data ingestion -- (a) directly from the Twitter Stream at its natural rate, and (b) streaming data from tweets stored in files, where we programmatically control the streaming rate to test the limits of our solution. In both cases, the period of observation and control was 8 hours. The average velocity of tweets from the direct stream was 4.9 tweets per second, and the maximum rate was 23.78 tweets per second. In the simulation, we multiplied the velocity up to 5 times with 5\% - 20\% duplicate tweets.   
\begin{figure*}[ht!]
	\centering
	\includegraphics[width=0.7\textwidth , scale=0.8]{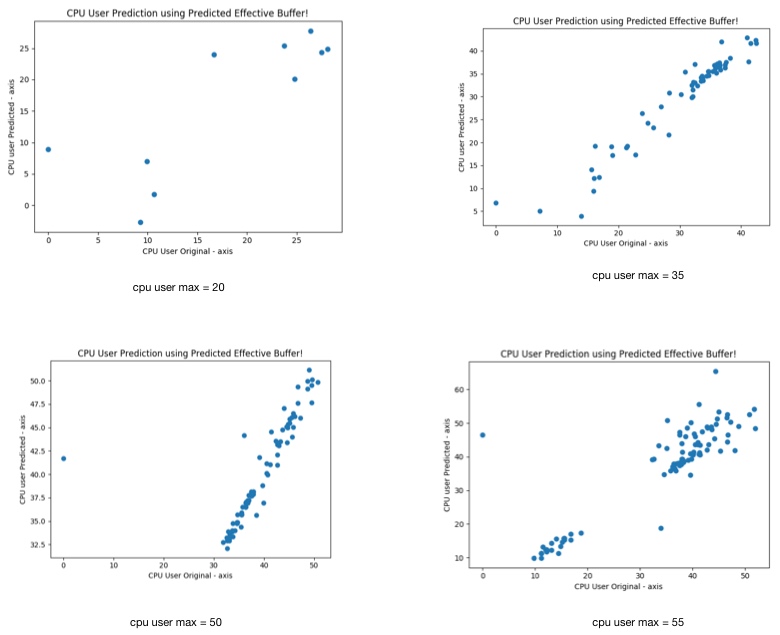}
	\caption{CPU User Predictions }
	\label{fig:cpu-pred}
\end{figure*}

\noindent \textbf{Implementation Architecture.} The stream processing architecture in our experimental setup (Figure \ref{fig:arch}) is designed as a producer-consumer model. Under the control of the ingestion controller module, the graph ingestor accepts data from the buffer and distributes it over multiple threads to construct the edge table concurrently. The results from the partial edge tables are collected in the Cypher statement buffer which performs the database commit operation. The ingestion controlled governs this process by using the performance monitoring services.
\subsection{Prediction Models}
We have tested two prediction models for $\beta_e$, the effective buffer size as function of the graph density $d$ and the bucket diversity ratio $\rho$ (Eq. \ref{eq2}) as well as $\mu_{exp}$ the expected CPU-usage as a function of $\beta_e$ (Eq. \ref{eq.4}). The model was tested with python SciKit learn\footnote{https://scikit-learn.org}.
\newline
\textbf{Expected Buffer Estimation($\beta_e$): }  
We determined that the equation $\beta_e[i] = K[i].\phi_1(\rho[i]) + R[i].\phi_2(d[i])$ is modeled with $\phi_1$ is linear while $\phi_2$ fits best with a quadratic function. The linear parameters K[i] and R[i] were estimated as 0.597 and 1.48, with standard error 0.024 and 0.021 respectively. 
\newline
\textbf{Expected CPU-Usage Estimation($\mu_{exp}$):}
Choosing an appropriate model for CPU-usage was a little more challenging. Table \ref{tab:regress} shows the models we have tested for and their errors. Our experiments $\mu_{exp}[n] = 0,09.log(\beta) + 0.01\mu_{exp}[n-1] + c$ is the closest fit while a linear model is a close second, Figure \ref{fig:cpu-pred} shows the observed values (X-axis) vs. the predicted values (Y-axis) the expected CPU-usage for 4 different settings of $cpu_{max}$. It is seen that a low choice of $cpu_{max}$ produces unclear results but the prediction closely matches the observation for $cpu_{max} > 30$. As $cpu_{max} > 50$, we observe that while the prediction is still good, the model demonstrates that the CPU-usage takes a quantum jump, explaining the gap in the plot in the CPU-usage range [0.21, 0.35].
\begin{figure*}[ht!]
\vspace{-1em}
	\centering
	\includegraphics[width=0.9\textwidth , scale=0.99]{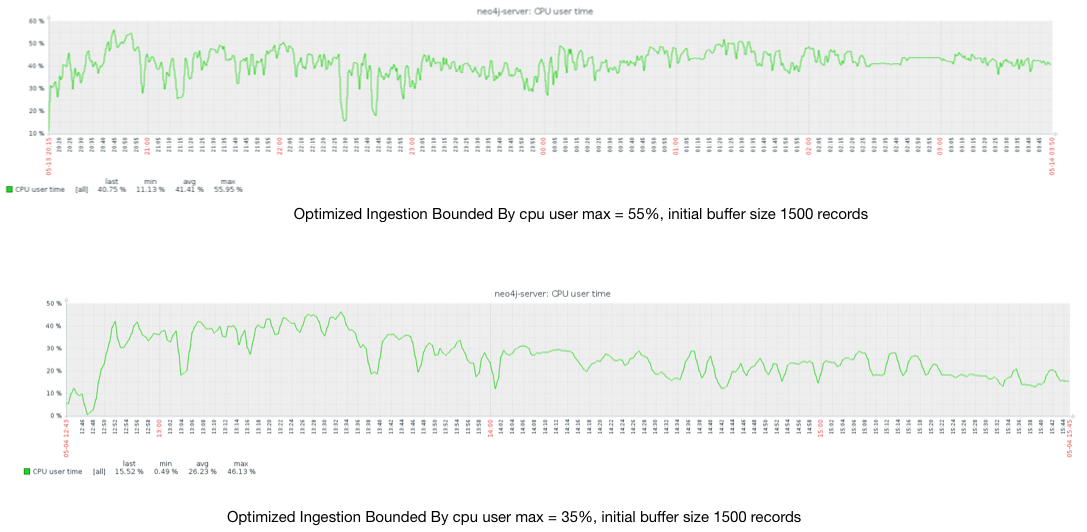}
	\caption{The CPU utilization is controlled as $cpu_{max}$ is set at 35\% and 55\% respectively. }
	\label{fig:sysm-opt-2}
\end{figure*}

\subsection{Effect of Graph Compression}
\begin{figure}[t]
	\centering
	\includegraphics[width=0.45\textwidth, scale=0.3]{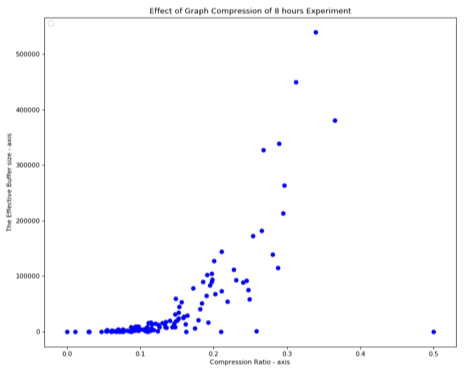}
	\caption{Effect of Graph Compression}
	\label{fig:effectofcompression-1}
\end{figure}

The experimental summary of the graph compression is shown in Figure \ref{fig:effectofcompression-1}: the X-axis represents the compression ratio (the effective count of insert instructions over the number of original tweets), and Y-axis represents the effective buffer size at that time. We can observe that for most cases, the effective compression rate (mean compression rate = 24.97\%) varies between 15\% and 35\% . With increased buffer size, the impact of the compression is not as effective. We have observed that during a twitter storm (e.g., in January 2018 for the hashtag \texttt{\#ReleasetheMemo}), when the graph density is high, the algorithm gives a better compression ratio. 

\subsection{Performance Output of The Algorithm}
\begin{figure}[ht!]
	\centering
	\includegraphics[width=0.45\textwidth , scale=0.9]{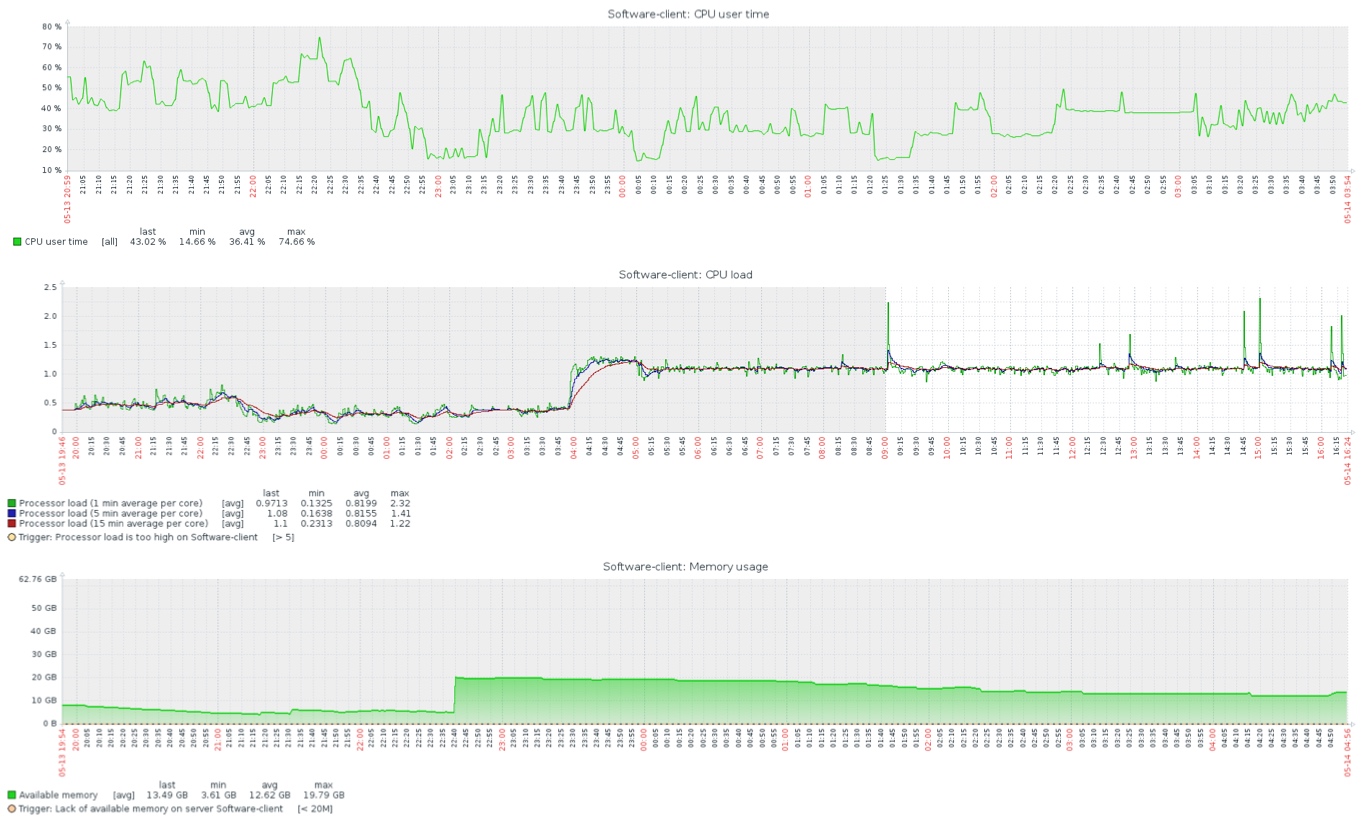}
	\caption{The CPU-usage, avg. CPU load, and memory use on the Ingestor Node stays within reasonable bounds for 8 hours.}
	\label{fig:client-perf}
\end{figure}

 The performance improvement of the final buffer is shown in Figure \ref{fig:sysm-opt-2}. 
One advantage using the political tweet data set is that the minimum data rate for tweets is high, and during peak hours there is a 4.5 fold increase in the data rate. In contrast with the uncontrolled CPU-usage, the experiment  (Figure \ref{fig:sysm-opt-2}) shows that after our technique is applied, the CPU user time never reaches a spiking condition. As expected, each time the CPU touches the maximum allowable limit, the algorithm reduces it down by not pushing any data during that period. Further, the IPS and context switching of the CPU was in low and stable condition, while the memory usage for the entire observation period is generally low. We monitored the system performance on the ingestor machine as well \ref{fig:client-perf}, and observed that CPU and memory utilization is well within control. 

\section{Conclusion}
\label{sec:conclusion}
In this paper, we addressed a bandwidth gap problem encountered in ingesting and storing social media data. We took advantage of the temporal clustering property of social media data (i.e., the fact that many similar nodes and edges are created during bursty periods) to compress the graph to save ingestion time, and dynamically adjusted to buffer to control the CPU load. Our work sits in the middle of graph analytics research underlying many data science applications \cite{ayman2019influence, pereira2016evolving, inoubli2018experimental} who use small data sets, and graph database research that promotes in-database graph analytics \cite{kronmueller2018graph} who do not consider streaming input. 
We view the graph stream ingestion problem discussed in this paper as a component of optimized ingestion control in the AWESOME polystore system \cite{gupta:bigdata:2016,DBLP:conf/bigdataconf/DasguptaMG17} where multiple streams of heterogeneous data can flow into a component DBMS managed under the polystore. We expect that the general idea of using buffer control, data compression and resource monitoring for DBMS can be effectively applied. In future work, we expect to extend this work to cover a larger variety of data models and data stores. 

Secondly, notice that in Algorithm \ref{alg:1}, we form a data structure that contains generic graph properties like degree distribution for the time-slice of data available in the buffer. Metrics like these are the building blocks of more complex analytical measures developed by graph-centric research communities. In future work, we will materialize more of these temporally evolving properties and use them for the evolutionary analysis of the social media graph, community detection \cite{ayman2019influence, pereira2016evolving, inoubli2018experimental}, and other graph analytics operations \cite{kronmueller2018graph}, which will benefit from our continuous computation of these ``building block'' measures.

Finally, our work primarily solves an infrastructure problem that generalizes beyond just social media data. As part of our future work, we will apply, and if needed, extend our system for other forms of structured and semi-structured streaming data (e.g., newswire data, lifelog data \cite{gurrin2014lifelogging}).

\bibliographystyle{abbrv}
\bibliography{escience}

\end{document}